\documentclass[12pt,preprint]{aastex}
\usepackage[usenames]{color}

\usepackage{rotating}

\shorttitle{Evolved variable stars in the Carina dSph}
\shortauthors{Coppola et al.}

\begin{document}

\title{The Carina Project. VI. The helium burning variable stars\altaffilmark{1}}

\author{
G. Coppola\altaffilmark{2},
P. B. Stetson\altaffilmark{3},
M. Marconi\altaffilmark{2}, 
G. Bono\altaffilmark{4,5},
V. Ripepi\altaffilmark{2},
M. Fabrizio\altaffilmark{6},
M. Dall'Ora\altaffilmark{2},
I. Musella\altaffilmark{2},
R. Buonanno\altaffilmark{4,7},
I. Ferraro\altaffilmark{5},
G. Fiorentino\altaffilmark{8},
G. Iannicola\altaffilmark{5},
M. Monelli\altaffilmark{9,10},
M. Nonino\altaffilmark{11},
L. Pulone\altaffilmark{5}, 
F. Th\'evenin\altaffilmark{12} and 
A. R. Walker\altaffilmark{13} 
}

\altaffiltext{1}{Based on images collected with the MOSAICII camera available at the CTIO 4m Blanco telescope, La Serena; (2003B-0051,2004B-0227, 2005B-0092, P.I.: A.R. Walker) and in part with the WFI available at the 2.2m MPG/ESO telescope (A064.L-0327) and with images obtained from the ESO/ST-ECF Science Archive Facility.} 
\altaffiltext{2}{INAF--Osservatorio Astronomico di Capodimonte, Via Moiariello 16, 
80131 Napoli, Italy; email: coppola@na.astro.it} 
\altaffiltext{3}{Dominion Astrophysical Observatory, NRC-Herzberg, 5071 West Saanich Road, Victoria, BC V9E 2E7, Canada}
\altaffiltext{4}{Dipartimento di Fisica-Universit\`{a} di Roma Tor Vergata, Via della Ricerca Scientifica 1, I-00133 Roma, Italy;}
\altaffiltext{5}{INAF-Osservatorio Astronomico di Roma, Via Frascati 33, I-00040 Monte Porzio Catone, Italy}
\altaffiltext{6}{INAF-Osservatorio Astronomico di Collurania, via M. Maggini -- 64100, Teramo, Italy}
\altaffiltext{7}{Agenzia Spaziale Italiana Science Data Center (ASDC), c/o ESRIN, via G. Galilei, I-00044 Frascati, Italy}
\altaffiltext{8}{INAF-Osservatorio Astronomico di Bologna, via Ranzani 1, 40127 Bologna, Italy}
\altaffiltext{9}{Instituto de Astrof\'{i}sica de Canarias, Calle Via Lactea s/n, 38205 La Laguna, Tenerife, Spain}
\altaffiltext{10}{Departamento de Astrof\'{i}sica, Universidad de La Laguna, 38200 Tenerife, Spain}
\altaffiltext{11}{INAF-Osservatorio Astronomico di Trieste, via G. B. Tiepolo 11, I-40131 Trieste, Italy}
\altaffiltext{12}{Universit\'{e} de Nice-Sophia Antipolis, Lab. Lagrange, UMR 7293, Observatoire de la C\^{o}te d’Azur, BP 4229, 06304 Nice, France}
\altaffiltext{13}{Cerro Tololo Inter-American Observatory, National Optical Astronomy Observatory, Casilla 603, La Serena, Chile}

\date{\centering drafted \today\ / Received / Accepted }

\begin{abstract}
We present new optical ($BVI$) time-series data for the evolved variable stars in the Carina 
dwarf spheroidal galaxy. The quality of the data and the 
observing strategy allowed us to identify 14 new variable stars. Eight out of the 14 are 
RR Lyrae (RRL) stars, four are Anomalous Cepheids (ACs) and two are geometrical variables. 
Comparison of the period distribution for the entire sample of RRLs with similar 
distributions in nearby dSphs and in the Large Magellanic 
Cloud indicates that the old stellar populations in these systems share similar properties. 
This finding is also supported by the RRL distribution in the Bailey diagram.  
On the other hand, the period distribution and the Bailey diagram of ACs display 
significant differences among the above stellar systems. This evidence suggests 
that the properties of intermediate-age stellar populations might be affected 
both by environmental effects and structural parameters.    
We use the $BV$ Period--Wesenheit (PW) relation of RRLs together with evolutionary 
prescriptions and find a true distance modulus of 
20.09$\pm$0.07(intrinsic)$\pm$0.1(statistical) mag that agrees quite well with similar 
estimates available in the literature.    
We identified four peculiar variables. Taking into account their position in the Bailey 
diagram and in the $BV$ PW relation, two of them (V14, V149) appear to be candidate 
ACs, while two (V158, V182) might be peculiar RRLs. In particular, the variable 
V158 has a period and a $V$-band amplitude very similar to the low-mass RRL 
---RRLR-02792---recently identified by~\citet{pietrzynski12} in the Galactic bulge.   
\end{abstract}


\maketitle

\section{Introduction} \label{sec:introduction}
Nearby dwarf spheroidal galaxies (dSphs) play a fundamental role in modern astrophysics. 
Wide-field imagers available at the 4--8m class telescopes provided a 
complete census of their stellar content down to the main sequence of 
the old stellar population~\citep{bono10,stetson11,monelli12}. 
Multi-slit and multi-fiber spectrographs available at the same 
facilities have provided the opportunity to investigate their kinematic 
structure~\citep{battaglia08,fabrizio11} and their 
metallicity distribution~\citep{clementini05,fabrizio12,lemasle12,venn12}. 
Dwarf spheroidals are also important laboratories for near-field
cosmology. Recent investigations indicate that the galaxies obey 
a well defined relation between mass and metallicity, where more massive 
galaxies are more metal-rich than less massive ones. This evidence 
has been typically explained as a consequence of galactic outflows, 
wherein more massive systems are able to retain metal-rich gas, while 
galactic outflows expel it in low-mass stellar systems. However, 
we are facing the evidence that nearby dSphs are, at fixed mass, 
more metal-poor than expected by the canonical mass-metallicity 
relation~\citep{chilingarian11}.  
        
Moreover, it has recently been shown 
that the metallicity---in stellar systems with total mass smaller than 
$10^{10}M_\odot$---is anti-correlated with the star-formation rate 
(SFR,~\citealt{mannucci10}). 
This empirical evidence is typically explained as a local infall of metal-poor 
gas that simultaneously increases the SFR and decreases the mean 
metallicity of these systems. Recent investigations have indicated 
quite complex star 
formation activity in nearby dSphs. For example it has been suggested that
Leo$\,$I experienced almost continuous star formation during the last 10~Gyrs~\citep{fiorentino12}, while Carina experienced multiple and distinct star 
formation events~\citep{bono10}. Moreover, several dSphs 
in the Local Group (LG) show broad metallicity distributions~\citep{hill10} 
suggesting that they have been able to retain the supernova yields 
of previous stellar generations.       
 
Variable stars in nearby dwarfs are important benchmarks for 
determining not only their distance and their geometry~\citep{minniti03,ripepi12,inno13},
but also their stellar
populations~\citep{pritzl05,pietrzynski06,kinemuchi08,kuehn08,szewczyk08,moretti09,musella09,szewczyk09,pietrzynski10,musella12,matsunaga11,dallora12}. 

Nearby dwarfs also offer the unique opportunity to study the 
dependence of stellar pulsation properties on the environment and, in 
particular, on the chemical enrichment history~\citep{caputo05,lanfranchi06}.
In this context the Carina dSph plays a key role, since it hosts 
a significant number of evolved and unevolved~\citep{mateo98} variable 
stars tracing the different episodes of star formation 
(see e.g.~\citealt{dallora03,monelli03,bono10}). 

In this paper we present an updated investigation of the variable 
star content in Carina based on new time-series data collected with 
the MOSAICII camera available at the CTIO 4m Blanco telescope and 
with the WFI camera available at the 2.2m MPG/ESO telescope. 
In Section~\ref{sec:sec2} we present the observations and the 
data-reduction strategy. In Section~\ref{sec:sec3} we discuss the 
location of new and old evolved variable stars in the color-magnitude 
diagram (CMD). In Section~\ref{sec:sec4} we compare the period distribution and the pulsation properties of evolved 
Carina variables with similar 
variable stars in nearby dwarfs. The conclusions and a few final 
remarks close the paper.

\section{Observations \& Data Reduction}\label{sec:sec2}
Our imaging data for the main body of Carina\footnote{We also have
observations of flanking fields to the south and northeast of the
galaxy, but we do not discuss these here.} are divided into 23
observing runs, as detailed in Table~\ref{tab:tab1}. To be more precise, there
were in fact 21 different observing runs, but in two cases (here
named wfi17/W00jan and B03jan/double) we obtained incomplete sets
of observations from the same observing run via two different
sources. In each case, after elimination of duplicate
observations, the non-redundant sets of images were kept as two
separate ``observing runs'' lest any inconsistencies in the
preprocessing of the different sets of images lead to subtle
changes in the absolute flux calibrations of the data.
The observations of Carina were contained within 327 distinct
datasets, where a dataset consists of either (a) all images
obtained by a given CCD on a given photometric night, or (b) all
images obtained by a given CCD on one or more non-photometric
nights during a given observing run.  

Table~\ref{tab:tab1} lists the number of different {\it
exposures\/} obtained of Carina, but because most of the
observations were made with mosaic cameras, the actual number of
{\it CCD images\/} that we processed was much greater (the
``Multiplex'' column in the table indicates the number of distinct
CCD images obtained in the course of each exposure).  
In total, we processed 5,946 distinct CCD images
for this photometric study of the main body of Carina.  
Processing of these data was carried out in our usual fashion with the \\
DAOPHOT\,IV/ALLSTAR/ALLFRAME suite of programs (e.g.~\citealt{stetson87,stetson94}). Of the 327 datasets
containing Carina observations, 223 were considered photometric
and were transformed individually to the photometric system
of~\citet{landolt92} via observations of Landolt's primary
photometric standards as well as secondary standards established
by~\citet{stetson00,stetson05}. The remaining 104 datasets were
considered non-photometric; for each of these the color-dependent
transformations to the Landolt photometric system were determined
from all available observations of primary and secondary
standards, but the photometric zero point of each individual CCD
image was determined by reference to secondary standards contained
within the image itself.  

Because of the non-overlapping nature of the fields of the different CCDs in
each mosaic camera, no individual star appears in more than a
small fraction of the total number of images. The {\it maximum\/}
number of flux measurements for any given star was 17 in $U$, 167
in $B$, 222 in $V$, and 85 in $I$. The 14 exposures in the $R$
filter (comprising 112 individual CCD images) were all obtained on
non-photometric occasions; as a result, Stetson was unable to
define local secondary standards in the $R$ photometric bandpass,
and we were not able to calibrate these $R$-band observations to the
Landolt photometric system. Nevertheless, these 112 $R$-band
images were included in the ALLFRAME reductions of Carina in order
to make use of the information they contribute to the completeness
of the star list and the precision of the astrometry.

\section{Variable Stars}\label{sec:sec3}
We applied a slightly more robust variant of the~\citet{welch93}
variability search technique to the new multi-band optical photometry of the main
body of Carina. On the basis of these data we have identified 
108 evolved variable stars; among them 14 are new identifications. A simple string-length
algorithm was applied to the time-series 
data to search for periodicity. Fourier series were then fitted by non-linear least-squares
to refine the periods and determine mean magnitudes and amplitudes.

For eight of the already known variables 
(six ACs: V27, V115, V149, V178, V187, V203 two RR Lyraes: V22, V176) 
we have updated the pulsation periods. Indeed, the previous search for 
variable stars had been performed on $B$,$V$ time-series data 
covering only three consecutive nights~\citep{dallora03}, and a few
period determinations
were affected by alias problems. Table~\ref{tab:prop} lists (from left to right) 
the identification, the classification, the epoch of maximum light, 
the period (days), the magnitude-averaged [($V$), ($B$)], the intensity-averaged 
[$\langle V \rangle$, $\langle B \rangle$], the $V$ and the $B$ amplitude 
(A$_{V}$, A$_{B}$) for the entire sample of variable stars identified in the 
current photometric survey. 

Eight out of the 14 new variables are RR Lyrae stars (RRLs), four are candidate 
anomalous Cepheids (ACs) and two are geometrical variables (eclipsing binary, W~UMa). 
The pulsation properties of the new variables are listed in Table~\ref{tab:tab2}, 
A more detailed discussion of the pulsation characteristics of the entire sample 
will be addressed in a future paper (Coppola et al. 2013, in preparation). 

\begin{figure}
\epsscale{1}
\plotone{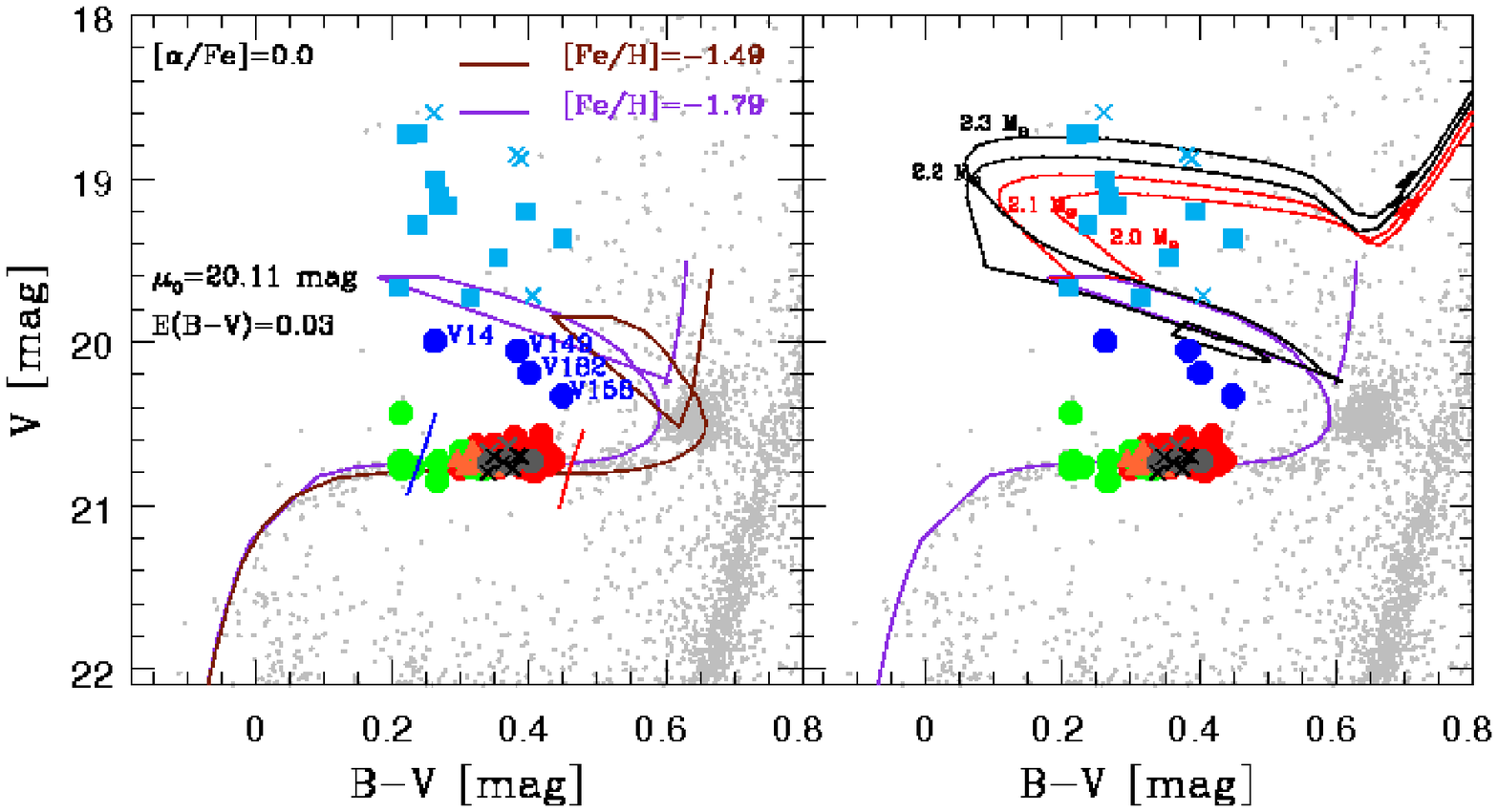}
\caption{Left: V, B-V Color--Magnitude Diagram (CMD) showing theoretical 
predictions for central helium-burning structures, based on scaled-solar 
evolutionary models constructed assuming a primordial helium of $Y=0.25$ and 
different iron abundances (see legend) from the BaSTI 
database~\citep{pietrinferni04}. Brown and purple solid lines represent 
the central helium-burning sequence for stars with stellar masses 
ranging from $\sim 0.5$ to $2.8$ $M_{\odot}$. Blue and red solid lines in the 
left panel delimit the theoretical instability strip for RRLs provided 
by~\citet{dicriscienzo04}. Red and green circles represent fundamental and 
first-overtone RRLs. Orange triangles and gray circles stand for 
double-mode pulsators and candidate Blazhko RRLs. Cyan symbols 
represent ACs. The blue circles stand for peculiar variables whose position 
in the CMD is intermediate between RRLs and ACs. Crosses indicate newly 
discovered variables.
Right: same as the left, but here the solid curves represent 
the evolutionary changes in the stellar properties near the transition mass
between central He burning in an electron degenerate core (red
lines) and quiescent central helium burning (black lines), and are
presented only for the more metal-poor abundance patterns. The
mass values are also labelled.}\label{fig:fig1}
\end{figure}

Figure~\ref{fig:fig1} shows the $V$,($B-V$) CMD of Carina 
in the region across the helium burning phases for both the old 
(horizontal branch [HB]) and the intermediate-age stellar population 
(red clump, [RC]). The 69 RRLs associated with the old stellar 
populations have been plotted as circles: red and green circles mark 
RRLs pulsating in the fundamental (RR$_{ab}$, 57) and in the first overtone 
(RR$_{c}$, 12) modes, respectively, while orange triangles mark the six 
candidate mixed mode variables, i.e., variables oscillating simultaneously 
in two or more pulsation modes (RR$_{d}$). 
The five RRLs showing modulation in both amplitude and phase 
(i.e., the so-called Blazhko RRLs) are plotted as grey circles. 
The fundamental RRLs newly detected have been plotted as crosses.  
The sixteen variables brighter than the RRLs have been classified 
according to~\citet{dallora03} as fundamental-mode ACs: 
cyan squares represent previously known variables while the cyan
crosses indicate, once again, our new AC candidates. Four
variables located between the RRLs and the ACs have been plotted
as blue circles, since the nature of these objects, based only on
their location in the CMD, is not yet clear.  

To further constrain the evolutionary status of the Carina variables, 
solid curves in the left panel of Figure~\ref{fig:fig1} represent the theoretical 
central helium-burning sequence for both old and intermediate-age stellar 
structures with the labeled solar--scaled chemical compositions
and stellar masses ranging from 0.5 to 2.8
$M_{\odot}$. The black and red curves in the right-hand panel of Figure~\ref{fig:fig1}
represent theoretical evolutionary tracks for stars near the transition
mass between central helium burning in degenerate (red) and non-degenerate
cores (black); these are shown, for clarity, only for the lower of the two metallicity values.
The adopted true distance modulus and reddening~\citep{dallora03} are also labelled.
The metal abundances we have adopted for this illustration follow current 
spectroscopic measurements based on high resolution spectra~\citep{fabrizio12,venn12,lemasle12}.  

\begin{figure}
\epsscale{1}
\plotone{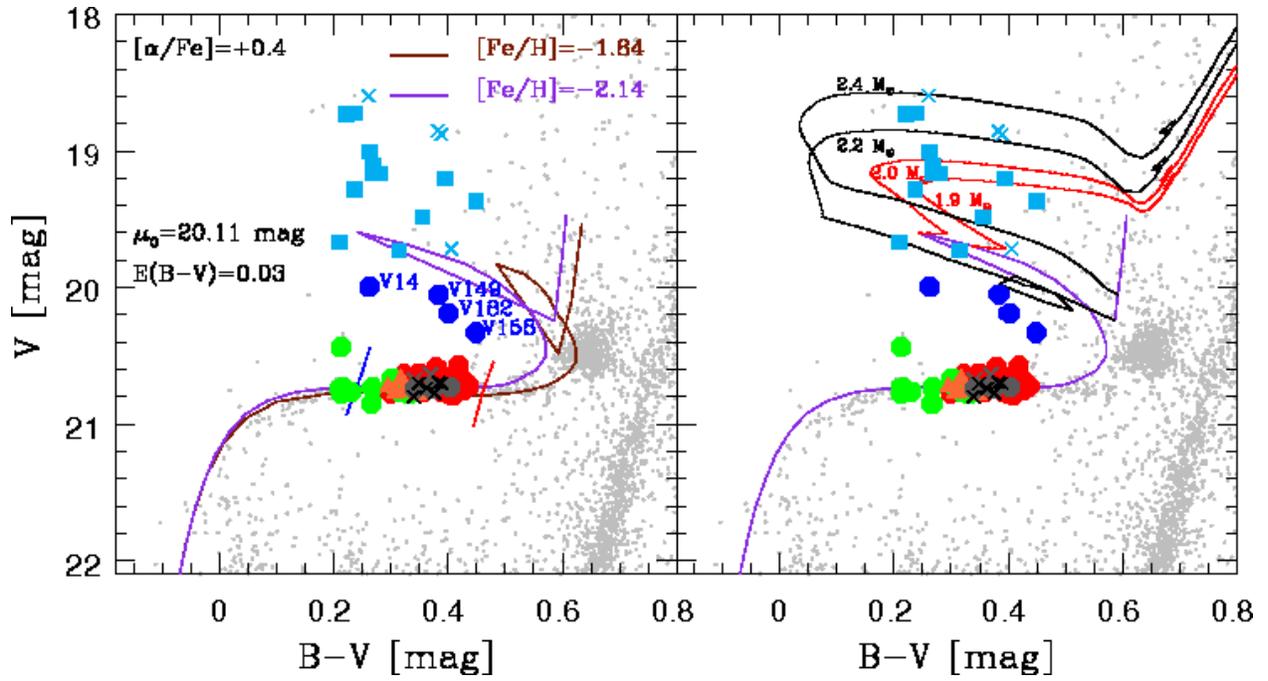}
\caption{Same as Figure~\ref{fig:fig1}, but for $\alpha$-enhanced helium burning 
evolutionary models.}\label{fig:fig2}
\end{figure}

The data plotted in this figure display several interesting features: 

\noindent a) The observed RRLs agree quite well with predicted ZAHBs. The same 
conclusion applies to the distribution of RRLs within the 
predicted instability strip. Indeed, only a few RR$_{c}$ variables appear 
to be slightly bluer than the predicted first-overtone blue edge 
for Z=0.0004 ([Fe/H]=--1.7 dex).\\
b) The ACs are only partially explained by current low-mass
evolutionary prescriptions, since a fraction of them ($V$$<$19.2 mag, ($B-V$)$<$0.3 mag)
are systematically hotter and brighter than predicted by partially 
electron-degenerate helium-burning models ($M/M_\odot<$ 2.2; red curves in the
right-hand panel of Figure~\ref{fig:fig1}).
The data plotted here support the suggestion that the brighter ACs 
are candidate short-period classical Cepheids. The difference between 
ACs and classical Cepheids is that the latter stars are slightly more 
massive ($M/M_\odot>$ 2.2 for the metal abundances considered here) 
and are characterized by quiescent helium-burning in non-degenerate cores.\\     
c) The intermediate-mass stars (AC, RC) show a spread in metallicity of the 
order of 0.25 dex as currently suggested by the spectroscopic measurements 
(\citet{fabrizio12} and references therein).\\ 
Figure~\ref{fig:fig2} shows the same data, but here the comparison is with 
alpha-enhanced ($[\alpha/$Fe$]=$~0.4) theoretical helium-burning structures. 
The data plotted in this figure indicate that a decrease in iron abundance to
[Fe/H]=--2.14 dex combined with the $\alpha$ enhancement (resulting in an overall 
metallicity of [M/H]=--1.79 dex, virtually the same as a solar--scaled abundance
pattern with [Fe/H]=--1.79 dex) has a minimal impact at least on the
RRLs. 
On the other hand, the blue circles in Figures~\ref{fig:fig1} and~\ref{fig:fig2} appear brighter than typical RRL, 
but fainter than ACs for the assumed metallicity. Therefore, the 
classification of these objects is more uncertain and further 
information is required to constrain their evolutionary and pulsational 
status.

\section{Pulsation properties}\label{sec:sec4}
\subsection{Period Distribution}\label{sec:sec4.1}
\begin{figure}
\epsscale{0.8}
\plotone{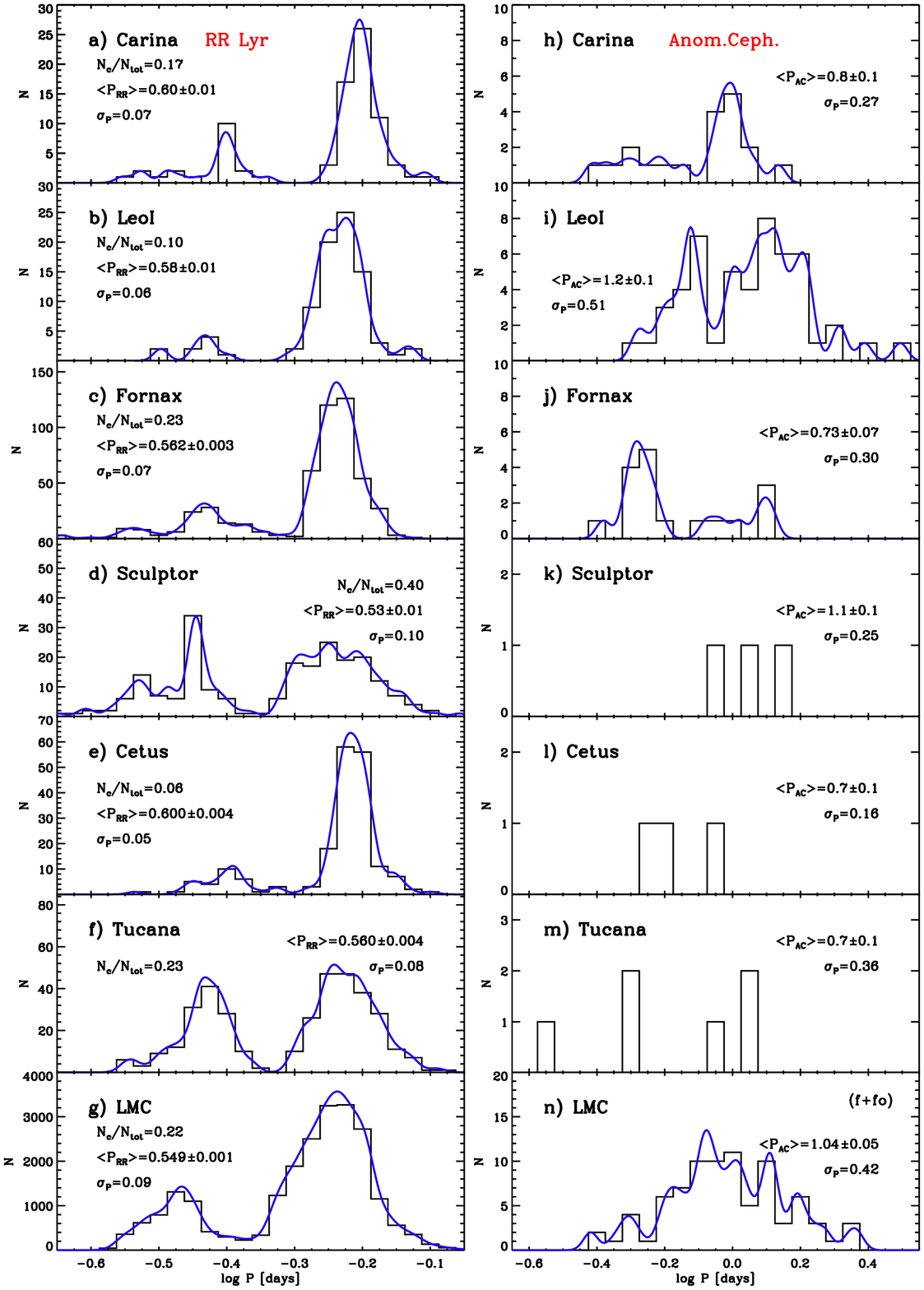}
\caption{\textit{Left}: from top to bottom we show the period 
distribution of the RRL samples ($RR_{ab}+RR_{c}+RR_{d}$) of Carina (82),
Leo$\,$I (95;~\citealt{fiorentino12}), Fornax (514;~\citealt{bersier02}), 
Sculptor (221;~\citealt{kaluzny95}), Cetus (155;~\citealt{bernard09}), Tucana (298;~\citealt{bernard09} and the 
LMC (95; OGLE-III sample by~\citealt{soszynski08,soszynski09}). 
\textit{Right}: same as the left, but for ACs.}\label{fig:fig3}
\end{figure}

The period distribution of the Carina RRLs is shown in the top left 
panel \textit{a)} of Figure~\ref{fig:fig3}. The period distribution 
shows two well-defined peaks for RR$_{ab}$ ($\log$ P $\sim$ --0.19) 
and RR$_{c}$ ($\log$ P $\sim$ --0.4) variables, plus a third minor 
peak at ($\log$ P $\sim$ --0.5). 

The mean period of fundamental RRLs is a crucial observable in the 
definition of the so-called Oosterhoff dichotomy~\citep{oosterhoff39}, 
typical of Galactic globular clusters (GGCs). 
The GGCs hosting RR Lyrae stars can be split, according 
to the mean period of RRL stars, into two different groups: 
the Oosterhoff type I [OoI], with $<$P$_{ab}$$>$ $\sim0.55$~days and the 
Oosterhoff type II [OoII] with $<$P$_{ab}$$>$ $\sim0.65$~days. 
In order to provide a more complete picture of the Carina variable star 
content, the periods of both first overtone and double-mode pulsators 
were fundamentalized according to the relation $\log{P_F}=\log{P_{FO}}+0.127$. 
The mean period of the entire sample of RRLs is then
$<$P$_{RR}$$>\;= 0.60 \pm 0.01$ days. This suggests that Carina 
can be classified as an OoII system, in agreement with previous 
results~\citep{dallora03}. However, the Oosterhoff dichotomy also shows up 
in the RRL population ratio, i.e., the ratio between the number of RR$_c$ and the total 
number of RRLs. 
The OoI clusters have a ratio $\sim 0.17$, while for OoII clusters 
this ratio is $\sim 0.44$~\citep{clement01}. On the basis of the new RRL sample, 
we found N$_c$/N$_{tot} \sim 0.17$, thus suggesting that this system is, 
according to the RRL population ratio, more similar to an 
OoI cluster. This discrepancy suggests that Carina is, in fact,
Oosterhoff-intermediate, like Draco~\citep{kinemuchi08}, Ursa Minor~\citep{nemec88}, 
Large Magellanic Cloud (LMC) globular clusters~\citep{bono94} and the outer-halo 
globular cluster Palomar~3~\citep{kinemuchi08}.

To better understand the period distribution of Carina variables we compare their properties with those of other LG dSphs. 
We start with the pulsating variables in the Leo$\,$I dSph ([Fe/H]=--1.43, 
$\sigma$=0.33 dex;~\citealt{kirby11}), including 86 RRLs and 51 ACs,  
have recently been investigated by~\citet{fiorentino12}. 
The period distribution of this galaxy's RRLs peaks at 
$<$P$_{RR}$$>\;= 0.58 \pm 0.01$ days (see panel \textit{b)} of Figure~\ref{fig:fig3}), 
thus also resembling an Oosterhoff-intermediate system. On the other hand, 
Leo$\,$I appears to be an OoI system according to the RRL population 
ratio --N$_c$/N$_{tot}$-- that is close to $\sim 0.10$. The indication is
that the pulsation properties of RRLs in Carina and in Leo$\,$I 
are generally similar. 

The period distributions of RRLs in the Fornax dSph 
([Fe/H]=--0.99, $\sigma$=0.36 dex;~\citealt{kirby11}; Figure~\ref{fig:fig3} 
panel \textit{c)}) are taken from~\citet{bersier02}. 
The value of $<$P$_{RR}$$>\;\sim 0.56$ days is also typical of an
Oo-intermediate system, while according to the RRL population
ratio (N$_{c}$/N$_{tot} \sim 0.23$) Fornax seems to resemble 
an OoI type globular cluster. We will discuss in more detail the 
variable star content of Fornax in Section~\ref{sec:sec4}.   

The period distribution of the 221 RRLs in the Sculptor dSph 
([Fe/H]=--1.68, $\sigma$=0.48 dex;~\citealt{kirby09,kirby11})  
by~\citet{kaluzny95} peaks around a mean period $<$P$_{RR}$$>\;= 0.53 \pm 0.01$ 
days (panel~\textit{d)} of Figure~\ref{fig:fig3}), suggesting an OoI system, 
but once again this is in contrast with the value of N$_c$/N$_{tot} \sim 0.40$ 
making it more similar to an OoII system. 

Together with the above satellite dwarf galaxies we decided to 
include in our analysis two isolated dwarfs, namely 
Cetus and Tucana~\citep{bernard09}. The former galaxy hosts 155 RRLs
and the period distribution shows a well defined main peak with a mean 
period $<$P$_{RR}$$>\;\sim 0.60$ days (panel~\textit{e)} of 
Figure~\ref{fig:fig3}), suggesting an OoII system. In spite of the large 
sample of RRLs, Cetus only hosts eight first overtones. This means that 
the RRL population ratio is quite small N$_{c}$/N$_{tot}$=0.06, therefore 
suggesting an OoI system. 
The period distribution of RRL (298) in Tucana is quite different 
when compared with Cetus. The mean period is $<$P$_{RR}$$>\sim 0.56$ 
days (panel~\textit{f)} of Figure~\ref{fig:fig3}) and suggests an OoI system.       
Moreover, Tucana hosts a sizable sample of first overtones (82) that is almost
the 30\% of the entire sample, and indeed the RRL population ratio 
attains a value significantly larger N$_{c}$/N$_{tot}$=0.23. This means that 
Tucana could be a ``pure'' OoI system. This and the evidence that this system hosts 
a high fraction of mixed-mode variables (60) makes Tucana a very interesting 
laboratory to investigate the occurrence of this still poorly understood 
pulsation phenomenon~\citep{bernard09}.  
The current results for Cetus and Tucana appear even more appealing if
we take into account the fact that they have a similar mean metallicity 
([Fe/H]$\sim$--1.8 dex) and a similar internal abundance dispersion 
($\sim$0.2 dex,~\citealt{bernard09}).   

Finally, according both to the value of $<$P$_{RR}$$>\;\sim 0.55$ days and to 
the fraction of RR$_{c}$, N$_c$/N$_{tot} \sim 0.22$, the LMC 
([Fe/H]=--1.48, $\sigma$=0.29 dex;~\citealt{gratton04}) resembles an 
OoI system~\citep[see Figure~\ref{fig:fig3} panel~\textit{g)}]{soszynski09}. 

The above empirical evidence brings forward a few interesting findings: 
the RRLs in nearby dSphs show similar period distributions, with 
pulsation properties ranging from OoI to OoII globulars, but the exact classification 
does depend on the diagnostic adopted to parameterize the pulsation properties. 
None of these dSph's can be described as a clean example of either OoI or OoII
according to {\it all\/} available classification criteria, with the only 
exception of Tucana, suggesting that the HB and the RGB luminosity function of 
this last system deserve a more detailed analysis.  
The RRLs in the LMC also show different properties, but this might be a consequence 
of the broad range in metal abundance~\citep{gratton04} and/or age shown by 
the old stellar component. The differences among the other dwarf galaxies and canonical
GGCs, considered together, are not yet clearly
understood.

A third peak at $\log$ P $\sim $--0.55 is present in all the above histograms,
possibly strengthening the inference that this feature is more likely an indicator
of a spread in metal abundance rather than a population of second-overtone RRLs, 
as originally suggested by~\citet{dallora03}. We note here that
current nonlinear pulsation models do not predict second
overtone pulsators~\citep[see][]{bono97c} among RR Lyrae stars. 

Panel \textit{h)} of Figure~\ref{fig:fig3} shows the period distribution 
of the ACs in Carina. The period distribution of these variables ranges 
from $\log{P} \sim -0.4$ to $\sim +0.2$ with a well defined peak at $\sim 0.0$ and a secondary group 
located in the short period range ($\log{P} \sim$ --0.2). To begin to interpret
the pulsation properties of ACs in Carina, we compared their period 
distributions with similar samples of ACs in Leo$\,$I, Fornax, and 
the LMC.
The period distribution of the ACs in Leo$\,$I~\citep[see Figure~\ref{fig:fig3}, panel~\textit{i)}]{fiorentino12} 
peaks at periods that are systematically longer than in Carina, 
and indeed the mean period is systematically longer 
(1.2$\pm$0.1 vs 0.8$\pm$0.1 days). Moreover, the ACs in Leo$\,$I also show a tail 
in the long-period range ($\log{P} \sim$ 0.2--0.6) that 
is not present in Carina.  

The period distribution of ACs in Fornax 
(panel \textit{j)} of Figure~\ref{fig:fig3}) is even more puzzling,
since it shows two well-defined peaks. However, in contrast with 
the Carina and Leo$\,$I samples, the main peak of the Fornax ACs 
is in the short-period range ($\log{P} \sim$--0.25). 
This is a peculiar feature, since the number of ACs in this 
period range is in the other dSphs small (see also the case 
of LMC in panel \textit{n)} of Figure~\ref{fig:fig3}).  

The number of ACs currently known in Sculptor, Cetus and Tucana does not 
allow us to reach any firm conclusion concerning their pulsation properties.  

The period distribution of the ACs in the LMC ranges from $\sim -0.4$ to 
$\sim 0.4$ (see panel~\textit{n)} of Figure~\ref{fig:fig3}) with a well defined 
mode at $\log{P} \sim$ --0.1. The period histogram is characterized by 
a broad distribution extending across both the short and the long period range.

The above findings indicate that the ACs in nearby dSphs display significant 
differences among these stellar systems. The same conclusion applies 
to ACs in the LMC, thus suggesting that the intermediate-age 
star formation and enrichment history of these systems followed 
different paths~\citep{fiorentinomonelli12}. 
In this context it is worth recalling that ACs in Leo$\,$I 
display a period distribution more skewed toward longer periods than 
ACs in the LMC. This evidence might be the consequence either of a 
very recent star formation episode or a systematic difference in metal 
abundance or both.

\subsection{The Period-Amplitude Diagram}\label{sec:sec4.2}

\begin{figure}
\epsscale{1}
\plotone{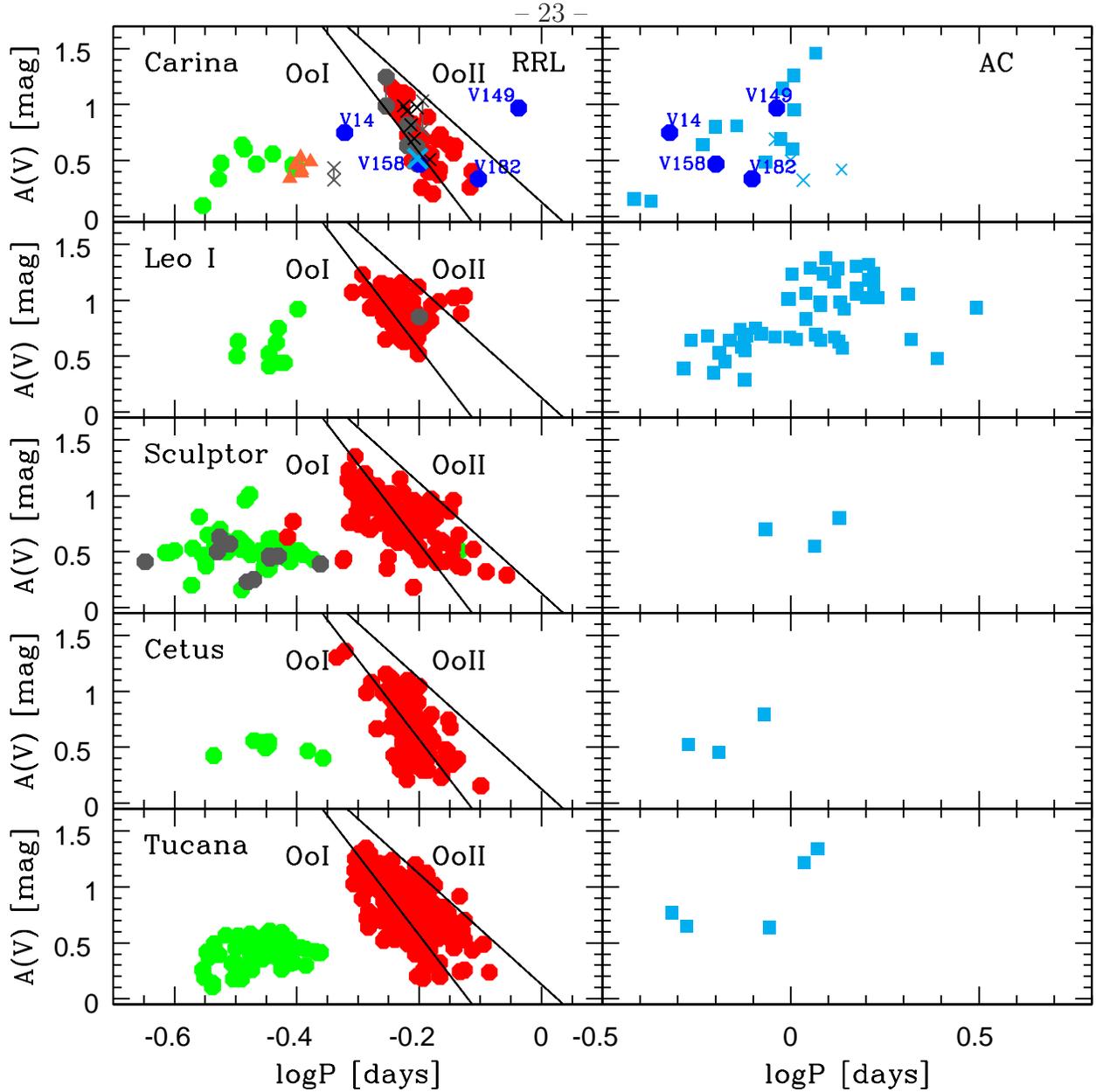}
\caption{Bailey diagram ($V$-band amplitude vs logarithmic period) for RRLs  
(left) and ACs (right) in Carina, Leo$\,$I, Sculptor, Cetus and Tucana. 
The solid lines display the positions of OoI and OoII Galactic globular 
clusters according to~\citet{clement00}. Symbols are the same as in 
Figure~\ref{fig:fig1}. The cyan cross in the left top panel marks the 
position of the peculiar RRLYR-02792 recently discovered 
by~\citet{pietrzynski12}.}\label{fig:fig4}
\end{figure}

To further investigate the pulsation properties of helium burning variable stars in Carina 
we also adopted the Bailey diagram, i.e., the luminosity amplitude vs logarithmic period. 
From top to bottom, the first two left panels of Figure~\ref{fig:fig4} show, once 
again, that the RRLs in Carina and in Leo$\,$I (Stetson et al. 2013, in preparation) have similar properties. 
Indeed, the RR$_{ab}$ variables display $V$-band 
amplitudes that are between the normal trend of OoI and OoII 
globular clusters (black solid lines). However, the RR$_c$ 
in Carina cover a slightly broader range in periods when 
compared with RR$_c$ in Leo$\,$I. 
It is worth mentioning that both the Blazhko and double-mode RRLs cover very limited 
ranges of period. The former group clusters around 
$\log$ P $\sim$ --0.25 to --0.20, while the latter 
is concentrated around $\log$ P $\sim$ --0.40. 
The comparison with Leo$\,$I is hampered by the fact 
that only two Blazhko RRLs are currently known in this system.     

Moving to the lower panels, the RR$_{ab}$ in Sculptor, 
Cetus and Tucana cover a broad range of both periods and amplitudes, 
but they seem to be in better agreement with the properties of OoI 
globulars. The RR$_{c}$ in Carina, Sculptor, Cetus and Tucana show 
similar behavior, displaying the so-called ``bell-shaped'' 
distribution, more typical of OoII globular clusters~\citep{bono97a}. 
However, the spread in amplitudes and periods of RR$_{c}$ in Carina 
and in Cetus is significantly smaller than in Sculptor and in Tucana. 
This empirical evidence further suggests that the spread in metallicity 
of the old stellar population in Carina and in Cetus is smaller 
than in Sculptor and in Tucana. The broad RR$_{c}$ period distribution 
of Sculptor and Tucana might also be affected by evolutionary 
effects~\citep[see their Figure~11]{bono97b}. 
The RR$_{ab}$ in Tucana display, at fixed period, a broad range in 
amplitudes. However, they appear to be more similar to OoII than to 
OoI type. This further support the evidence that Tucana is mainly 
an OoI system.
Finally, we mention that, unlike those in Carina, the Blazhko RRLs in 
Sculptor are RR$_{c}$ variables and cover a broad range in period.  

The data plotted in the right panels indicate, once again, that ACs in Carina and Leo$\,$I 
show quite different properties. In the latter system a majority ($\sim$ 60\%) of 
ACs have $V$-band amplitudes larger than 0.9 mag, while in the firts only five out of the 
18 ACs ($\sim$ 28\%) have such large amplitudes. The current analysis is hampered by the fact 
that we still lack a solid diagnostic to discriminate between fundamental and 
first-overtone ACs. Carina contains
two short-period ACs with small $V$-band amplitudes (0.1--0.2 mag), 
but their position in the Period-Wesenheit diagram does not appear 
to be peculiar (see below).   

It is worth noticing that, according to the Bailey diagram, the
four peculiar variables (V14, V149, V158 and V182, indicated by blue circles) 
appear to be candidate ACs, and indeed
they attain amplitudes similar to the other ACs in the same period range.  
However, two (V158, V182) out of the four might be candidate 
RRL stars. The reason is twofold: a) they are located in the same region 
of the Bailey diagram as the other RRLs, while the variables V14 and 
V149 exhibit amplitudes that are too large for their periods; 
b) they are both fainter and redder than the other ACs (cf. Figure~\ref{fig:fig2}). 
The hypothesis that V158 and V182 are candidate RRL stars was first
suggested by~\citet{monelli03}. In particular, these authors suggested
that V158 and V182 might be the aftermath of 
intermediate-mass stars that during their evolution experienced violent mass-loss 
events. The existence of this class of objects has been soundly demonstrated 
by~\citet{pietrzynski12}. 
They identified an RR Lyrae-like variable in an eclipsing binary system (RRLYR-02792) 
located in the Galactic bulge and provided a firm estimate of its dynamical mass.
They found that its mass is 0.26 $M/M_\odot$, thus confirming its peculiar nature. 
This object mimics a typical RRL, but its evolutionary status is significantly 
different. The main energy source of this low-mass variable seems to be 
hydrogen-shell burning, since current theoretical and empirical constraints indicate 
that central helium burning---typical of RRLs---can take place only in structures more massive 
than $\approx$~0.5$M/M_\odot$~\citep{castellani07}. The peculiar evolutionary history 
of RRLYR-02792 is also supported by the fact it is characterized by a large negative 
period derivative (--8.4$\pm$2.6$\times$ 10$^{-6}$ days/year), thus further 
supporting the difference with canonical RRLs for which the same derivative is 
typically two orders of magnitude smaller~\citep{kunder11}. 
Interestingly enough, the prototype of this new class of variable stars in the Bailey 
diagram is located (cyan cross) very close to V158, i.e., one of the two peculiar
candidate RRLs. Unfortunately, the current data do not allow us to estimate the 
period derivatives of V158 and V182. 

\begin{figure}
\epsscale{0.6}
\plotone{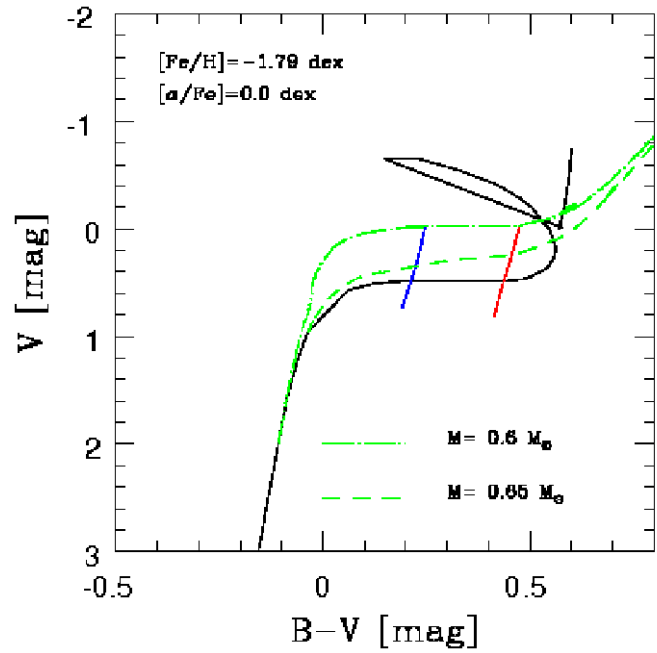}
\caption{Predicted V, B-V Color-Magnitude Diagram of the scaled--solar 
metal--poor helium burning sequence plotted in Figure~\ref{fig:fig1}. The two green lines 
display the evolution of two hot HB stellar structures, while the blue and 
the red almost vertical lines display the edges of the RR instability strip.}\label{fig:fig6}
\end{figure}

To further investigate the nature of the two peculiar RRL candidates, 
we considered whether they might 
be either evolved RRL or candidate type II Cepheids (P2C). To our knowledge P2C 
variables have been identified only in Fornax by~\citet{bersier02}, but they 
might also be present in Carina. Figure~\ref{fig:fig6} shows the predicted $V$,($B-V$) CMD of the 
scaled-solar metal-poor helium burning sequence plotted in Figure~\ref{fig:fig1}, together with 
the blue edge for first overtone pulsators and the red edge for fundamental pulsators. 
Using the pulsation relation provided by~\citet{dicriscienzo04}, we found that the 
periods at the intersection between the ZAHB and the instability strip are 
$P_F^{FO}$=0.279 and $P_F$=0.787 days, where the former has been fundamentalized.      
This period range agrees quite well with the observed range of Carina RRLs. 
To further investigate the possibility either of evolved RRLs or P2C we took into 
account the off-ZAHB evolution of two old HB stellar structures with stellar 
masses smaller than the typical masses of RRLs (0.70 $\le$$M/M_\odot$$\le$ 0.75). 
The two green lines cross, as expected, the instability strip at higher luminosity, 
and therefore they produce RRLs with longer periods. The structure with $M/M_\odot$=0.65 
at the intersection between the ZAHB and the instability strip gives periods 
of $P_F^{FO}$=0.371 and $P_F$=1.131 days, while the structure with $M/M_\odot$=0.60 
gives periods of $P_F^{FO}$=0.558 and $P_F$=1.530 days. These findings further 
indicate that the period of the two peculiar RRLs (V158, 0.632 days; V182, 0.778 days) are 
too short compared with the predicted ones. Indeed pulsation and evolutionary 
prescriptions indicate that pulsators located close to red edge of the 
instability strip and at least one-half magnitude brighter than typical RRLs should 
have periods longer than one day, i.e., they should be P2C of the BL
Herculis type~\citep{marconi07, marconi11}.    

\subsection{The Period-Wesenheit relation}\label{sec:sec4.3}

\begin{figure}
\epsscale{1}
\plotone{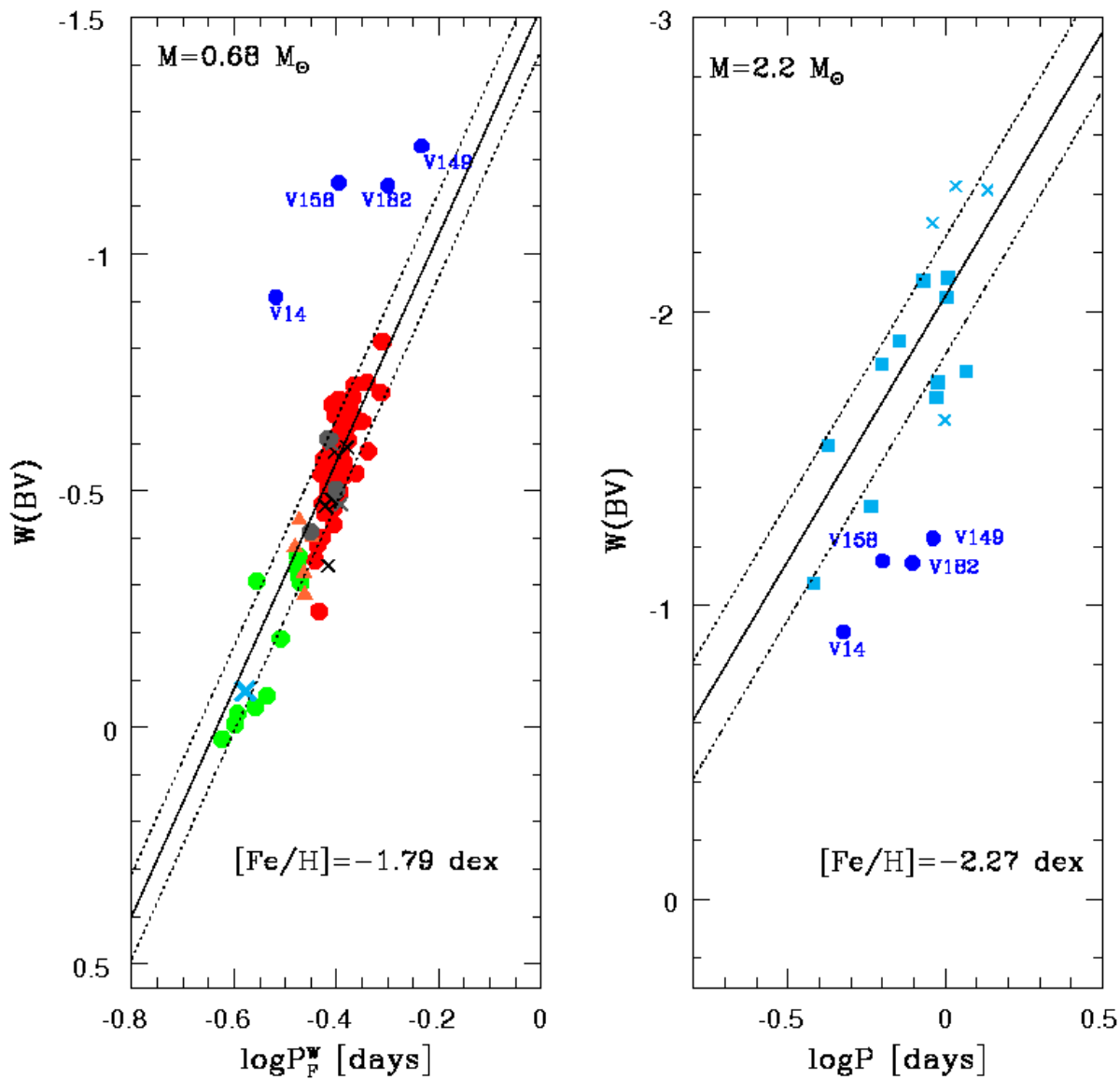}
\caption{$BV$ Period--Wesenheit relations for RRLs (left) and ACs (right) 
in Carina. Symbols are the same as in Figure~\ref{fig:fig1}. The solid lines 
show the predicted behavior at constant mass and metallicity according to 
pulsational models for RRLs by~\citet{dicriscienzo04} and for ACs 
by~\citet{marconi04}. The dotted lines depict the intrinsic dispersion 
of the above relations.}\label{fig:fig5}
\end{figure}

Additional insights about the evolutionary status of the four peculiar variables, 
can be obtained adopting the Period-Wesenheit (PW) relation. The main advantage in using 
Wesenheit magnitudes 
is that they are reddening-free by definition, being estimated using both apparent 
magnitudes and colors linked by a coefficient given by an extinction
law~\citep[see e.g.][and references therein]{marconi04}. Note that we do not expect,
a priori, that foreground or internal reddening in Carina will be a serious issue for our work; at Galactic
longitude and latitude $(260^\circ,-22^\circ)$ the foreground extinction is expected to
be small and indeed we have previously adopted E(B-V)~=~0.03 mag for this direction (cf. Figure~\ref{fig:fig2}), and
no interstellar material has yet been identified within Carina. A secondary advantage
of the Wesenheit magnitude is that it also largely removes the dependence of period
on temperature at fixed luminosity, resulting from the finite width of the instability strip:
loci of constant period on the observational CMD are nearly parallel to the reddening vector~\citep[see e.g.][]{stetson98}. The Wesenheit magnitude therefore produces
a narrower period-apparent magnitude relation one based on simple $B$- or $V$-band magnitudes.
The data plotted in Figure~\ref{fig:fig5} show the $BV$ PW 
relations for Carina RRLs (left panel) and ACs (right panel). 
The solid lines show the predicted PW relations at 
constant mass and metallicity according to pulsational models 
for RRLs computed by~\citet{dicriscienzo04} and for ACs 
computed by~\citet{marconi04}. The dotted lines display the 
intrinsic dispersion of the above relations. The left panel 
shows the Wesenheit magnitude versus the $\log$ P$_{F}^{W}$. This parameter 
depends on the period (RRc variables were fundamentalized), the mass 
and the metallicity according to the following relation: 

\begin{equation}
 \log P_{F}^{W} = \log P_{F} + 0.54 \log M/M_{\odot} +0.03 \log Z,
\end{equation}

\noindent where the symbols have their usual meaning (see for more details~\citealt{dicriscienzo04}).
Adopting the iron abundance and the mean mass indicated in left panel of Figure~\ref{fig:fig5}, we 
found a true distance modulus of $\mu_{0}$=20.09$\pm$0.07(intrinsic)$\pm$0.1 (systematic) mag. 
The intrinsic error estimate accounts for uncertainties in the mean RRL magnitudes, the 
photometric zero points, and the intrinsic dispersion of the theoretical PW diagram. 
The systematic errors account for uncertainties in the
pulsational models. 
The peculiar nature of the above variables is further indicated by the fact 
that they exhibit Wesenheit magnitudes that are, at fixed period, systematically brighter 
than typical RRLs. The difference is at the 3$\sigma$ level, on average.

The empirical scenario becomes even more puzzling if we consider the PW relations
of ACs. Two (V14, V158) out of the four peculiar variables have Wesenheit 
magnitudes relatively close to the AC PW relation for the stellar mass 
and chemical composition indicated~\citep{marconi04}, while the other two 
are on average 6$\sigma$ fainter.   
However, the spread in magnitude of the AC PW relation is significantly larger than 
for RRL (0.2 vs 0.09 mag), thus suggesting that this is not a robust diagnostic 
of the evolutionary status of intermediate-age helium-burning variable stars. 
This evidence is further supported by the fact that the candidate short-period 
classical Cepheids (brighter ACs) seem to obey the same PW relation. 
The width of the observed AC PW relation is mainly due to a dispersion 
in mass and partially to evolutionary effects and mode identification.  

\begin{figure}
\epsscale{1}
\plotone{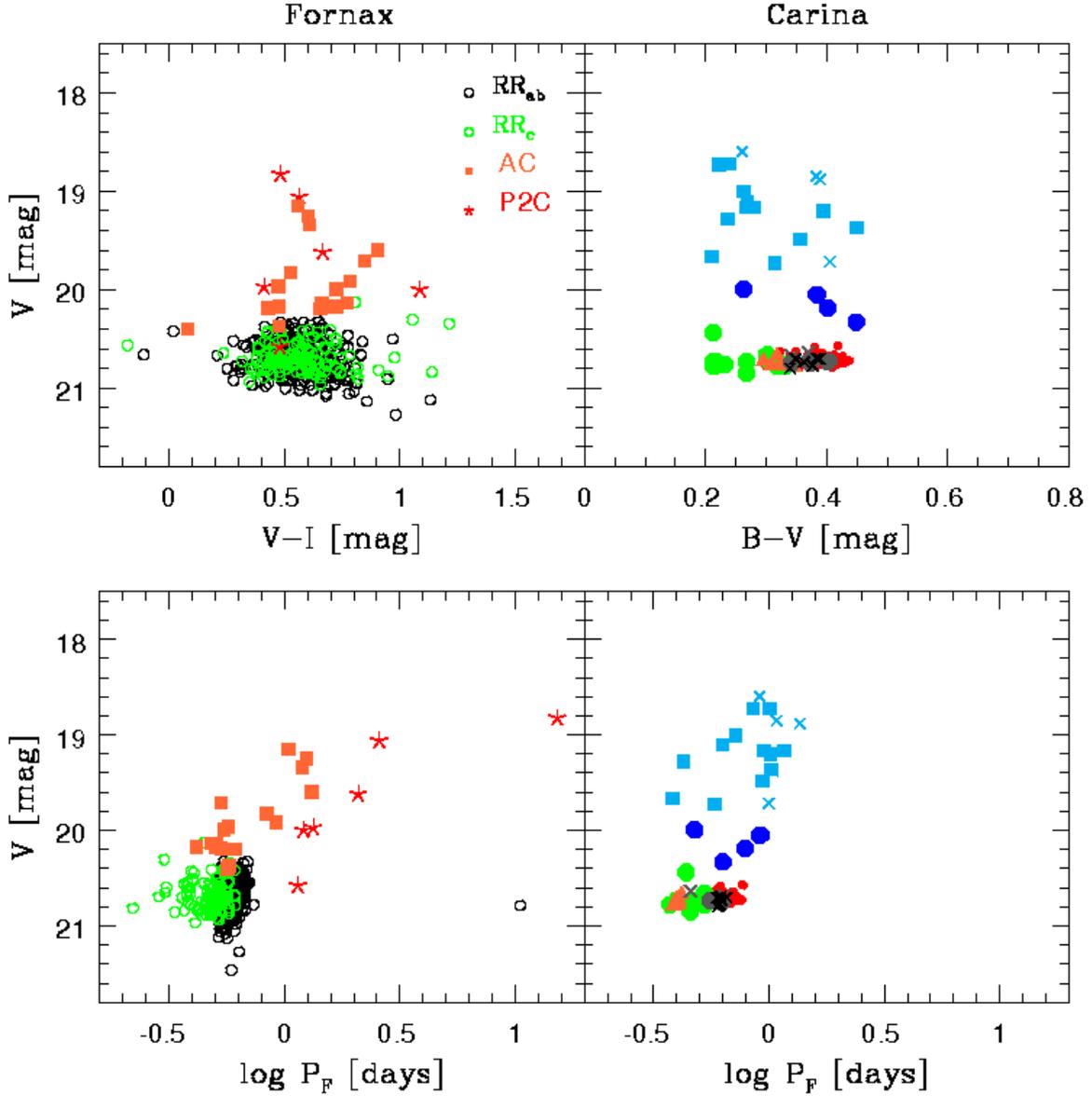}
\caption{Top left: V, V-I Color-Magnitude Diagram (CMD) of Fornax evolved variable 
stars. The RRLs, ACs and P2Cs were plotted using different symbols. 
Their apparent magnitudes were rescaled to Carina assuming for Fornax a true 
distance modulus of $\mu_{0}$=20.65 mag and a reddening of E(B-V)=0.025 mag~\citep{bersier00, bersier02}. 
The periods of FOs were fundamentalized. 
Top right: Same as the left, but for Carina evolved variable stars and in the 
V, B-V CMD.
Bottom left: Same as the top left, but in V-magnitude--logarithmic period plane.
Bottom right: Same as the left, but for Carina Variable stars.}\label{fig:fig7}
\end{figure}

Finally, we decided to compare in more detail the Carina evolved variable 
stars with their counterparts in Fornax, even if the census of evolved variable stars in this system is still far from being complete~\citep{bersier02}. To this end, the apparent magnitudes 
of RRLs in Fornax were rescaled to the 
apparent magnitude of RRLs in Carina assuming a true distance modulus of 
$\mu_{0}$=20.62 mag and a reddening of E(B-V)=0.025 mag~\citep{bersier00,bersier02}.    
The data plotted in the top left panel of Figure~\ref{fig:fig7} show that the distribution 
of evolved variables in Fornax is significantly different from that
observed in Carina. Canonical RRLs in Carina are separated from canonical 
ACs by almost one magnitude, while in Fornax there is an almost continuous 
transition between RRLs, ACs and P2Cs. The lack of a clear separation between 
RRL and ACs is suspicious, since evolutionary models predict a steady 
increase in luminosity when moving from the ZAHB to the intermediate-age  
helium burning sequence and a minimum gap of $\approx 0.8$ mag between the ZAHB and the evolved
variable magnitude level is expected even in the most metal poor
regime~\citep[see Figure~7 in][for details]{caputo04}. However, a spread in metallicity could cause
a spread in visual magnitude, and in turn smear out the separation 
between RRLs and ACs. 

To further investigate this interesting point, we performed the same comparison
but in the V-magnitude logarithmic period plane. Data plotted in the bottom 
panels show that Carina RRLs and ACs are well separated. On the other hand, 
the Fornax ACs split into two different groups: the short period group 
($\log P\le$--0.1) overlaps with RRLs, while the long-period group 
($\log P>$--0.1) is, at fixed period, brighter than the P2Cs. 
The long-period group has a canonical behavior, since ACs are approximately 
a factor of three more massive than than P2C and, at fixed period, they 
should be brighter.      
The short-period group appears peculiar, since their periods (0.44--0.56 days) 
cover the same period range as the evolved RRLs (see the discussion concerning the 
nature of the two peculiar RRL candidates in Carina). This preliminary evidence, if supported by new 
and more detailed investigations on Fornax evolved variables, might explain 
the peculiar peak in the period distribution of Fornax ACs.       

\section{Conclusions and final remarks}
We have presented a new census and analysis of helium-burning variable stars 
in the Carina dSph. Their observed properties have been compared with 
theoretical predictions to constrain their evolutionary and pulsational status and their 
distance. The main results of our analysis are the following:
 
\begin{itemize}
 \item[i)] we have identified eight new RRLs that are found to share the same general
properties as the whole sample. In particular, they agree quite well 
with the predicted ZAHB and instability strip for a metallicity ranging 
from [Fe/H]=--1.79 to --1.49 dex. The RRL period distribution shows a 
remarkable similarity with the RRLs in Leo$\,$I, Fornax and the LMC.
Using the theoretical $BV$ PW relation for the metallicity and the 
stellar mass inferred from the comparison with the theoretical ZAHB, 
we found a true distance modulus $\mu_{0}$=20.09$\pm$0.07$\pm$0.1 mag 
that agrees quite well with previous estimates in the literature~\citep{pietrzynski03,pietrzynski09}.

 \item[ii)] We have identified four new ACs with periods around one 
day. The comparison with evolutionary predictions suggests that the 
stellar mass of these objects ranges from $\sim$2.0 to 
$\sim$2.4 M$_{\odot}$. The current empirical evidence indicates that the 
bright tail of the distribution might be short-period classical Cepheids. This means 
that Carina and Leo$\,$I~\citep{fiorentino12} are good laboratories 
to study the transition from intermediate-mass stars characterized 
by quiescent central helium burning (M/M$_\odot$$>$2.2), 
producing classical Cepheids, to those burning helium in an 
electron-degenerate core (M/M$_\odot$$<$2.2), producing ACs. 
We also found that the period distribution of ACs is quite different 
among nearby dwarf galaxies. This occurence might indicate that the star formation 
history in the last few Gyrs differs strongly from system to system.  
This finding supports recent results by~\citet{fiorentino12}. 
 \item[iii)] We have investigated the properties of four already 
known variables that appear to be peculiar in the CMD, since their 
mean magnitudes are intermediate between RRLs and ACs. The comparison 
between predicted and observed periods indicates that they cannot be 
of the BL$\,$Her type, i.e., low-mass 
(M/M$_\odot$$\approx$0.50 to 0.60) HB stars evolving from the 
blue (hot) to the red (cool) region of the CMD and crossing the 
instability strip at luminosities brighter than typical RRLs. 
We found that their periods are 30\% shorter than predicted 
by pulsation models (typically in the range from 0.7 to 1.3 days). 
According to the Bailey diagram two (V14, V149) out of the four 
appear to be candidate ACs, whereas the variables V158 and V182 
might be peculiar RRLs as already suggested by~\citet{monelli03}. 
It is also interesting to note that in the Bailey 
diagram the variable V158 is located very close to the prototype 
---RRLYR-02792---of a new group of variable stars recently discovered 
by~\citet{pietrzynski12} and investigated by~\citet{smolec13}. 
This new group of variables mimics the properties of typical RRLs, 
but they have a mass that is a factor of two smaller. This suggests
that they are intermediate-mass stars that have experienced violent 
mass-loss events. 

 \item[iv)] A firm quantitative analysis of evolved variable stars in 
nearby dwarfs requires not only homogenous and accurate multiband 
photometry, but also time series data that cover a broad time interval.    
Only these data can open the path to a thorough spectroscopic investigation 
that can allow us to investigate how the environment, the chemical 
composition and the star formation history affect their evolutionary 
and pulsation properties. 
\end{itemize}

The above findings further emphasize the key role that evolved variable 
stars in dSphs can play to constrain the evolutionary and pulsation properties 
of low- and intermediate-mass stars. The similarity of the old stellar 
populations traced by RRLs in nearby stellar systems indicates that the 
early star formation in these systems was quite homogeneous.  
On the other hand, the difference in the intermediate-age populations, as
traced by ACs, suggests that recent star formation events in these systems 
does differ strongly from system to system.  

It goes without saying that detailed investigations of the pulsational 
properties of evolved variables in nearby isolated systems can shed new 
light on their individual properties and on their environmental influences.

\bigskip 

\acknowledgments   
This work was partially supported by PRIN--INAF 2011 ``Tracing the formation
and evolution of the Galactic halo with VST'' (PI: M. Marconi) and by
PRIN--MIUR (2010LY5N2T) ``Chemical and dynamical evolution of the Milky Way
and Local Group galaxies'' (PI: F. Matteucci). One of us (G.B.) thanks ESO for support as a science visitor.
Matteo Monelli was supported by the Education and Science Ministry of
Spain (grants AYA2010-16717).\\
It is a real pleasure to thank an anonymous referee for his/her very
positive opinion concerning the content of our paper and for his/her
valuable comments.


\clearpage

\clearpage

\begin{table}
\scriptsize
\caption[]{Log of observations.}\label{tab:tab1}
\begin{tabular}{lllllcccccc}
 \hline
 \hline
\noalign{\smallskip}
 & Run ID & Dates & Telescope &  Camera & $U$ & $B$ & $V$ & $R$ & $I$ & Multiplex \\
\noalign{\smallskip}
\hline
\noalign{\smallskip}
 1 & ct92:     &  1992 12 19-22         & CTIO 1.5m    & Tek2K-1  & -- & 30 & -- & -- & 42 &               \\
 2 & wfi3:     &  1999 03 17            & MPI/ESO 2.2m & WFI      & -- & -- &  1 & -- & -- &  $\times 8$   \\
 3 & fors9912: &  1999 12 02            & ESO VLT 8.0m & FORS1    & -- & -- &  1 & -- &  1 &               \\
 4 & wfi15:    &  1999 12 15-19         & MPI/ESO 2.2m & WFI      & -- & -- & 52 & -- & 16 &  $\times 8$   \\
 5 & B00jan:   &  1999 12 30-2000 01 10 & CTIO 4.0m    & Mosaic2  & -- & 48 & 48 & -- & -- &  $\times 8$   \\
 6 & wfi17:    &  1999 12 31-2000 01 07 & MPI/ESO 2.2m & WFI      & -- & 45 & 48 & -- &  4 &  $\times 8$    \\
 7 & W00jan:   &  2000 01 06-08         & MPI/ESO 2.2m & WFI      & -- & 26 & 28 & -- & -- &  $\times 8$    \\
 8 & wfi22:    &  2000 02 26            & MPI/ESO 2.2m & WFI      & -- & 13 & 12 & -- & 12 &  $\times 8$    \\
 9 & wfi21:    &  2000 03 05            & MPI/ESO 2.2m & WFI      & -- &  2 &  2 & -- &  2 &  $\times 8$    \\
10 & wfi14:    &  2000 10 29-01         & MPI/ESO 2.2m & WFI      &  5 &  8 &  8 & -- & -- &  $\times 8$    \\
11 & cg:       &  2001 01 17-18         & CTIO 4.0m    & Mosaic2  & -- & -- &  6 & -- & 12 &  $\times 8$    \\
12 & B02oct:   &  2002 10 31            & CTIO 4.0m    & Mosaic2  & -- &  5 &  5 & -- & -- &  $\times 8$    \\
13 & B02nov:   &  2002 11 29            & CTIO 4.0m    & Mosaic2  & -- &  5 &  5 & -- & -- &  $\times 8$    \\
14 & B03jan:   &  2003 01 02            & CTIO 4.0m    & Mosaic2  & -- &  4 &  3 & -- & -- &  $\times 8$    \\
15 & double:   &  2003 01 02            & CTIO 4.0m    & Mosaic2  & -- & -- &  1 & -- & -- &  $\times 8$    \\
16 & wfi19:    &  2003 03 05-07         & MPI/ESO 2.2m & WFI      & -- & 19 & 15 & -- & -- &  $\times 8$    \\
17 & B03octa:  &  2003 10 27            & CTIO 4.0m    & Mosaic2  & -- &  5 &  7 & -- & -- &  $\times 7$    \\
18 & B03nov:   &  2003 11 28            & CTIO 4.0m    & Mosaic2  & -- &  8 & 10 & -- & -- &  $\times 7$    \\
19 & B04jan:   &  2004 01 22-23         & CTIO 4.0m    & Mosaic2  & -- & 13 & 15 & -- & -- &  $\times 7$    \\
20 & B04jan29: &  2004 01 29            & CTIO 4.0m    & Mosaic2  & -- &  1 &  1 & -- & -- &  $\times 7$    \\
21 & B04dec11: &  2004 12 12            & CTIO 4.0m    & Mosaic2  &  4 & -- &  5 & -- &  7 &  $\times 8$    \\
22 & B04dec19: &  2004 12 20            & CTIO 4.0m    & Mosaic2  &  6 & -- & -- & -- & 12 &  $\times 8$    \\
23 & wfi29:    &  2008 09 28-10 07      & MPI/ESO 2.2m & WFI      & -- & 14 & 14 & 14 & 14 &  $\times 8$    \\
\hline
\end{tabular}

Notes:\\
 1  Observers: T.~Smecker-Hane, P.~B.~Stetson; \\
 2  Program ID: unknown, observer unknown;\\
 3  Program ID: 64.N-0421(A);\\
 4  Program ID: 000.H-0597;\\
 5  Observers: A.~Walker, C.~Smith;\\
 6  Program IDs: 064.N-0512, 0164.N-0210,064.L-0327;\\
 7  Program ID: a064.L-0327; proprietary data not found in the archive;\\
 8  Program ID: 164.O-0561(E), observer M.~Schirmer;\\
 9  Program ID: 064.N-0564, observer V.~Testa;\\
 10  Program ID: 164.O-0089(A), observer V.~Ripepi;\\
 11  Observers: C.~Gallart, J.~P.~Garcia;\\
 12  Proposal ID: 2002B-0077, observer A.~Walker;\\
 13  Proposal ID: 2002B-0077, observer A.~Walker;\\
 14  Observer: A.~Walker;\\
 15  Observer: A.~Walker, single exposure missing from Run \#14;\\
 16  Program ID: 70.B-0635(A); \#3 non-functional;\\
 18  Observer: A.~Walker; chip \#3 non-functional;\\
 19  Proposal ID: 2004b-051, observer A.~Walker; chip \#3 non-functional;\\
 20  Observers: C.~Aguilera, C.~Smith, A.~Walker; chip \#3 non-functional;\\
 21  Observer: A.~Walker;\\
 22  Observers: A.~Walker, M.~Monelli;\\
 23  Program ID: 081.A-9026(A).
\end{table}

\begin{deluxetable}{lllccccccc}
\tablewidth{0pt}
\tabletypesize{\footnotesize}
\tablecaption{Pulsation Properties of Variable Stars.}
\tablehead{ID & Type & Period &  (V)$^{a}$ & $\langle V \rangle^{b}$ & (B)$^{a}$ & $\langle B \rangle^{b}$ & A$_{V}$ & A$_{B}$\\
   ~~~~&   ~~~~ & (days) & (mag)	& (mag) & (mag)& (mag)	&(mag)& (mag)}
\startdata
V7   &      RR$_{ab}$   & 0.603314	& 20.766&  20.715&  21.159&  21.081&  1.082 &	1.265	 &	\\
V10  &      RR$_{ab}$ 	& 0.58451541	& 20.765&  20.712&  21.168&  21.085&  1.070 &   1.311	 &      \\
V11  &      RR$_{d}$    & 0.405505      & 20.780&  20.770&  21.095&  21.079&  0.413 &   0.498	 &     	\\
V14  &      AC	        & 0.4766970     & 20.027&  19.997&  20.303&  20.260&  0.749 &   0.886	 &      \\
V17  &      EB	        & 0.3933399     & 19.604&  19.603&  20.968&  20.967&  0.157 &   0.184	 &      \\
V22  &      RR$_{ab}$   & 0.6380664     & 20.769&  20.755&  21.180&  21.156&  0.569 &   0.706	 &      \\
V24  &      RR$_{ab}$   & 0.6182002     & 20.738&  20.712&  21.138&  21.096&  0.729 &   0.914	 &      \\   
V26  & 	    RR$_{d}$  	& 0.419218	& 20.684&  20.673&  21.008&  20.991&  0.488 &   0.532	 &      \\ 
V27  &      AC	        & 1.020387     	& 19.409&  19.369&  19.887&  19.819&  0.950 &   1.221	 &      \\   
V29  &      AC	        & 0.7178805     & 19.043&  19.008&  19.323&  19.271&  0.809 &   1.005	 &      \\   
V30  &      RR$_{ab}$ 	& 0.618813 	& 20.798&  20.774&  21.210&  21.173&  0.713 &   0.890	 &      \\   	    
V31  &      RR$_{ab}$ 	& 0.6457939     & 20.765&  20.754&  21.166&  21.149&  0.473 &   0.584	 &     	\\   	      
V33  &      AC	        & 0.5836253     & 19.753&  19.731&  20.077&  20.046&  0.639 &   0.773	 &      \\   
V34  &      RR$_{ab}$ 	& 0.5869487	& 20.804&  20.758&  21.174&  21.103&  1.079 &   1.302	 &      \\   	         
V40  &      RR$_{c}$    & 0.3926230     & 20.796&  20.784&  21.133&  21.116&  0.446 &   0.544	 &      \\
V43  &      RR$_{c}$    & 0.2992490     & 20.748&  20.735&  20.971&  20.952&  0.476 &   0.572	 &   	\\	     	    
V47  &      RR$_{c}$    & 0.3237693     & 20.786&  20.764&  21.026&  20.995&  0.639 &   0.735	 &   	\\	   
V49  &      RR$_{ab}$ 	& 0.6815054	& 20.706&  20.696&  21.118&  21.102&  0.419 &   0.535	 &      \\
V57  &      RR$_{ab}$   & 0.6129004     & 20.809&  20.787&  21.233&  21.198&  0.676 &   0.834	 &   	\\

V60  &      RR$_{ab}$   & 0.6094138     & 20.787&  20.756&  21.194&  21.147&  0.833 &   1.013	 &   	\\

V61  &      RR$_{ab}$ 	& 0.6213529	& 20.709&  20.682&  21.129&  21.086&  0.749 &   0.910	 &   	\\   
V65  &      RR$_{ab}$ 	& 0.6517114     & 20.754&  20.719&  21.150&  21.094&  0.888 &   1.116	 &   	\\
V67  &      RR$_{ab}$ 	& 0.603718      & 20.762&  20.735&  21.150&  21.104&  0.725 &   0.941	 &      \\ 
V68  &      RR$_{ab}$ 	& 0.678732      & 20.729&  20.722&  21.169&  21.159&  0.369 &   0.407	 &      \\
V73  &      RR$_{ab}$ 	& 0.5695186	& 20.819&  20.761&  21.174&  21.091&  1.151 &   1.307	 &      \\
V74  &      RR$_{c}$	& 0.3982350	& 20.783&  20.771&  21.106&  21.089&  0.439 &   0.534	 &   	\\
V77  &	    RR$_{ab}^{c}$ 		& 0.60431       & 20.754&  20.731&  21.178&  21.134&  0.730 &   0.869	 &      \\      
V84  &      RR$_{ab}$   & 0.616665     	& 20.755&  20.726&  21.119&  21.080&  0.827 &   0.901	 &      \\   
V85  &      RR$_{ab}$ 	& 0.640514     	& 20.677&  20.654&  21.081&  21.047&  0.689 &   0.828	 &   	\\
V87  &      AC	        & 0.855616      & 18.742&  18.728&  18.987&  18.967&  0.483 &   0.584	 &      \\       
V89  &      RR$_{d}$	& 0.387544      & 20.790&  20.785&  21.145&  21.132&  0.338 &   0.461	 &      \\ 	 
V90  & 	    RR$_{ab}$	& 0.631361 	& 20.816&  20.796&  21.239&  21.205&  0.625 &   0.786	 &      \\
V91  &      RR$_{ab}$ 	& 0.7180700     & 20.697&  20.683&  21.133&  21.109&  0.566 &   0.726	 &      \\
V92  &      RR$_{ab}$ 	& 0.6301265     & 20.767&  20.749&  21.205&  21.177&  0.627 &   0.761	 &      \\
V105 &      RR$_{ab}$ 	& 0.6323024	& 20.710&  20.690&  21.139&  21.107&  0.681 &   0.801	 &      \\
V115 &      AC		& 1.010975      & 18.751&  18.731&  18.980&  18.953&  0.601 &   0.678	 &      \\
V116 &      RR$_{ab}$ 	& 0.6833130     & 20.786&  20.763&  21.192&  21.153&  0.728 &   0.920	 &      \\
V122 &      RR$_{ab}$ 	& 0.6314708     & 20.695&  20.672&  21.096&  21.059&  0.719 &   0.873	 &      \\

V123 &      RR$_{ab}$ 	& 0.674967      & 20.690&  20.670&  21.114&  21.082&  0.687 &   0.778	 &   	\\
V124 &      RR$_{ab}$ 	& 0.5917212     & 20.709&  20.672&  21.099&  21.042&  0.884 &   1.069	 &      \\    
V125 &      RR$_{ab}$ 	& 0.5940966     & 20.693&  20.640&  21.044&  20.963&  1.104 &   1.326	 &      \\    
V126 &	    RR$_{ab}^{c}$		& 0.5570978     & 20.787&  20.742&  21.130&  21.085&  1.034 &   0.959	 &      \\
V127 &      RR$_{ab}^{c}$	  	& 0.626010	& 20.760&  20.747&  21.146&  21.121&  0.472 &   0.704	 &     	\\
V129 &	    AC		& 0.6301751	& 19.141&  19.105&  19.428&  19.374&  0.799 &   0.988	 &      \\             
V135 &      RR$_{ab}$ 	& 0.5909249	& 20.675&  20.630&  21.050&  20.979&  0.969 &   1.233	 &      \\
V136 &	    RR$_{ab}$	& 0.631613      & 20.708&  20.691&  21.127&  21.100&  0.580 &	0.728	 &      \\
V138 &      RR$_{ab}$   & 0.6392584     & 20.720&  20.704&  21.133&  21.109&  0.554 &	0.670	 &      \\   
V141 &	    RR$_{ab}$	& 0.6353331     & 20.754&  20.740&  21.170&  21.147&  0.528 &   0.685	 &      \\
V142 &      RR$_{c}$    & 0.3635433     & 20.752&  20.734&  21.031&  21.002&  0.559 &   0.696 	 &      \\ 
V143 &      RR$_{ab}$ 	& 0.6095792     & 20.704&  20.678&  21.100&  21.058&  0.756 &   0.945	 &      \\ 
V144 &      RR$_{c}$    & 0.3933563     & 20.677&  20.665&  20.985&  20.967&  0.431 &   0.556	 &      \\ 
V148 &      RR$_{c}$    & 0.326654      & 20.460&  20.441&  20.680&  20.654&  0.600 &   0.692	 &      \\ 
V149 &      AC	        & 0.917713      & 20.092&  20.053&  20.498&  20.437&  0.969 &   1.199	 &      \\ 
V151 &      RR$_{c}$    & 0.3418011     & 20.866&  20.852&  21.140&  21.119&  0.468 &   0.584	 &      \\ 
V153 &      RR$_{ab}$ 	& 0.6603690	& 20.701&  20.687&  21.099&  21.075&  0.552 &   0.681	 &      \\ 
V158 &      RR$_{ab}^{f}$ 	& 0.632464      & 20.342&  20.332&  20.800&  20.781&  0.468 &   0.630	 &      \\
V159 &      RR$_{ab}$ 	& 0.5751536     & 20.752&  20.707&  21.099&  21.030&  0.957 &   1.188	 &      \\ 
V164 &      RR$_{ab}$ 	& 0.633919      & 20.745&  20.731&  21.125&  21.098&  0.564 &   0.723    &      \\ 
V173 &	    RGB$^{d}$		& 0.660738	& 18.525&  18.524&  19.574&  19.573&  0.134 &   0.143    &	\\
V174 &      RR$_{ab}$ 	& 0.6531168     & 20.768&  20.759&  21.189&  21.175&  0.397 &   0.498	 &   	\\ 
V175 &      RR$_{c}$    & 0.3923768	& 20.760&  20.749&  21.089&  21.072&  0.461 &   0.539	 &   	\\ 
V176 &      RR$_{ab}$    & 0.764565	& 20.742&  20.741&  21.172&  21.169&  0.119 &   0.207    &      \\ 
V178 &      AC	        & 1.0155652     & 19.283&  19.204&  19.735&  19.599&  1.261 &   1.657	 &      \\ 
V179 &      RR$_{ab}$ 	& 0.663799      & 20.737&  20.735&  21.165&  21.162&  0.201 &   0.240	 &   	\\ 
V180 &      AC	        & 0.519033      & 	&  	 &  	  &  	   &  	    &   	 & 	\\
V181 &      RR$_{c}$    & 0.2794913     & 20.780&  20.779&  20.995&  20.993&  0.097 &   0.152	 &   	\\ 
V182 &      RR$_{ab}^{f}$ 	& 0.788970      & 20.198&  20.192&  20.603&  20.594&  0.337 &   0.411	 &   	\\ 
V183 &      RR$_{ab}$ 	& 0.6140441     & 20.599&  20.587&  20.987&  20.967&  0.497 &   0.630    &      \\ 
V184 &      RR$_{c}$    & 0.3951280     & 20.723&  20.713&  21.033&  21.016&  0.448 &   0.535	 &      \\ 
V185 &      RR$_{ab}$ 	& 0.620919      & 20.742&  20.720&  21.137&  21.101&  0.661 &   0.876	 &     	\\ 
V186 &      RR$_{ab}$ 	& 0.5790123	& 20.828&  20.775&  21.160&  21.075&  1.092 &   1.319	 &      \\
V187 &      AC	        & 0.950289	& 19.230&  19.164&  19.520&  19.432&  1.143 &   1.355	 &      \\
V188 &      RR$_{ab}$ 	& 0.5973782     & 20.726&  20.676&  21.092&  21.016&  1.087 &   1.330	 &      \\             
V189 &      RR$_{ab}$ 	& 0.7023692     & 20.662&  20.644&  21.058&  21.031&  0.643 &   0.761	 &      \\
V190 &      AC	        & 1.164708     	& 19.255&  19.166&  19.555&  19.447&  1.459 &   1.480	 &      \\
V191 &      RR$_{ab}$ 	& 0.6502902     & 20.703&  20.687&  21.109&  21.083&  0.566 &   0.714	 &      \\              
V192 &      RR$_{d}$    & 0.405096      & 20.737&  20.728&  21.045&  21.024&  0.387 &   0.629	 &      \\            
V193 &      AC	        & 0.4263580	& 19.283&  19.282&  19.521&  19.519&  0.134 &   0.212	 &      \\
V194 &	    FV$^{e}$		& 0.2645829     & 16.068&  16.067&  15.873&  15.872&  0.145 &   0.112	 &	\\
V195 &      RR$_{ab}$ 	& 0.6189680     & 20.782&  20.769&  21.146&  21.126&  0.500 &   0.611	 &      \\
V196 &      RR$_{ab}$ 	& 0.6674832     & 20.685&  20.677&  21.087&  21.073&  0.435 &   0.498	 &      \\
V197 &      RR$_{c}$    & 0.2961719	& 20.747&  20.741&  20.962&  20.953&  0.335 &   0.386	 &      \\   
V198 &      RR$_{d}$    & 0.395523	& 20.737&  20.725&  21.088&  21.071&  0.458 &   0.547	 &      \\   
V199 &      RR$_{ab}$ 	& 0.769925      & 20.582&  20.575&  21.005&  20.994&  0.401 &   0.483	 &      \\
V200 &      RR$_{ab}$ 	& 0.637343      & 20.749&  20.745&  21.169&  21.163&  0.257 &   0.327	 &      \\
V201 &      RR$_{ab}$ 	& 0.7222442     & 20.756&  20.738&  21.162&  21.135&  0.631 &   0.774	 &      \\       
V202 &      RR$_{ab}$ 	& 0.6150525	& 20.744&  20.722&  21.136&  21.097&  0.723 &   0.873	 &      \\             
V203 &      AC	        & 0.9398922	& 19.508&  19.487&  19.895&  19.843&  0.695 &   1.049	 &      \\             
V204 &      RR$_{ab}$ 	& 0.6332666	& 20.748&  20.731&  21.145&  21.121&  0.582 &   0.683	 &      \\
V205 &      AC	        & 0.3833677	& 19.668&  19.667&  19.879&  19.877&  0.158 &   0.188	 &      \\             
V206 &      RR$_{ab}$ 	& 0.5887446	& 20.757&  20.699&  21.149&  21.048&  1.097 &   1.444	 &     	\\
V207 &      RR$_{d}$    & 0.403913	& 20.778&  20.765&  21.106&  21.088&  0.528 &   0.598	 &      \\   

\enddata
\tablecomments{\\
$^{a}$ Magnitude-averaged magnitudes;\\
$^{b}$ Intensity-averaged magnitudes;\\
$^{c}$ Blazhko candidate;\\
$^{d}$ Variable located along the RGB;\\
$^{e}$ Field variable;\\
$^{f}$ Peculiar RR Lyrae.}\label{tab:prop}
\end{deluxetable}

\begin{table}
 \scriptsize
\caption[]{Pulsation Properties of new variable stars.}\label{tab:tab2}
\begin{tabular}{lllccccccc}
 \hline
 \hline
\noalign{\smallskip}
ID & Type & Period &  (V)$^{a}$ & $\langle V \rangle^{b}$ & (B)$^{a}$ & $\langle B \rangle^{b}$ & A$_{V}$ & A$_{B}$\\
   ~~~~&   ~~~~ & (days) & (mag)	& (mag) & (mag)& (mag)	&(mag)& (mag)\\
\noalign{\smallskip}
\hline
\noalign{\smallskip}
V208 &      RR$_{ab}$   & 0.656588 	& 20.716&  20.703&  21.112&  21.091&  0.508 &   0.624	 &      \\   
V209 &      RR$_{ab}$   & 0.6123133     & 20.806&  20.777&  21.202&  21.154&  0.816 &   1.012	 &      \\   
V210 &      RR$_{ab}^{c}$	        & 0.457815	& 20.647&  20.639&  21.023&  21.009&  0.378 &   0.497	 &      \\   
V211 &      RR$_{ab}$   & 0.6194511	& 20.721&  20.699&  21.119&  21.083&  0.697 &   0.859	 &     	\\   
V212 &      RR$_{ab}$   & 0.6264287     & 20.782&  20.740&  21.168&  21.103&  0.974 &   1.186    &      \\           
V213 &      RR$_{ab}$   & 0.5956516     & 20.745&  20.705&  21.119&  21.054&  0.987 &   1.241	 &      \\   
V214 &      RR$_{ab}^{c}$ 	& 0.639186     	& 20.708&  20.679&  21.086&  21.021&  0.774 &   1.148	 &      \\
V215 &      RR$_{ab}$   & 0.603031	& 20.844&  20.803&  21.217&  21.143&  0.941 &   1.196	 &      \\                
V216 &      AC	        & 1.079685      & 18.858&  18.852&  19.246&  19.235&  0.324 &   0.525	 &      \\    
V217 &      AC	        & 0.910935	& 18.625&  18.599&  18.899&  18.860&  0.689 &   0.853	 &      \\
V218 &      AC	        & 0.9986934	& 19.734&  19.719&  20.153&  20.125&  0.504 &   0.680	 &   	\\
V219 &      AC	        & 1.365582	& 18.892&  18.884&  19.287&  19.273&  0.420 &   0.551	 &      \\
V220 &      EB	        & 0.18632063	& 14.698&  14.674&  15.368&  15.343&  0.747 &   0.803	 &      \\
V221 &      WUma	& 1.785305	& 19.234&  19.233&  20.559&  20.558&  0.138 &   0.136	 &      \\    
\hline
\end{tabular}

$^{a}$ Magnitude-averaged magnitudes;\\
$^{b}$ Intensity-averaged magnitudes;\\
$^{c}$ Blazhko candidate.
\end{table}

\end{document}